\documentclass[a4paper,aps,prd,twocolumn,preprintnumbers,showpacs,superscriptaddress,nofootinbib]{revtex4}
\usepackage{graphics,graphicx,epsfig}
\usepackage{amsfonts,amsmath,amssymb}
\usepackage{amsmath}
\usepackage{amsfonts}
\usepackage{amssymb}
\usepackage{color}
\usepackage{graphicx}

\definecolor{darkgreen}{rgb}{0,0.35,0}

\begin{document}

\title{From Lorentz-Chern-Simons to Massive Gravity in 2+1 Dimensions}
\author{Sim\'on del Pino}
\email{simon.delpino.m-at-mail.pucv.cl}
\affiliation{Instituto de F\'{\i}sica, Pontificia Universidad Cat\'{o}lica de Valpara%
\'{\i}so, Casilla 4059, Valpara\'{\i}so, Chile.}
\author{Gaston Giribet}
\email{gaston-at-df.uba.ar}
\affiliation{Physique Th\'eorique et Math\'ematique, Universit\'e Libre de Bruxelles and International Solvay Institutes,
Campus Plaine C.P. 231, B-1050 Bruxelles, Belgium.}
\affiliation{Departamento de F\'{\i}sica, Universidad de Buenos Aires and IFIBA-CONICET,
Ciudad Universitaria, Pabell\'on I, 1428, Buenos Aires, Argentina.}
\affiliation{Instituto de F\'{\i}sica, Pontificia Universidad Cat\'{o}lica de Valpara%
\'{\i}so, Casilla 4059, Valpara\'{\i}so, Chile.}
\author{Adolfo Toloza}
\email{atoloza-at-cecs.cl}
\affiliation{Instituto de F\'{\i}sica, Pontificia Universidad Cat\'{o}lica de Valpara%
\'{\i}so, Casilla 4059, Valpara\'{\i}so, Chile.}
\affiliation{Centro de Estudios Cient\'{\i}ficos CECs Casilla 1469, Valdivia, Chile}
\author{Jorge Zanelli}
\email{z-at-cecs.cl}
\affiliation{Centro de Estudios Cient\'{\i}ficos CECs Casilla 1469, Valdivia, Chile}
\pacs{42}

\begin{abstract}
We propose a generalization of Chiral Gravity, which follows from considering a Chern-Simons action for the spin connection with
anti-symmetric contorsion. The theory corresponds to Topologically Massive Gravity at the chiral point non-minimally coupled to an additional scalar mode that gathers the torsion degree of freedom. In this setup, the effective cosmological constant (the inverse of the curvature radius of maximally symmetric solutions) is either negative or zero, and it enters as an integration constant associated to the value of the contorsion at infinity. We explain how this is not in conflict with the Zamolodchikov's $c$-theorem holding in the dual boundary theory. In fact, we conjecture that the theory formulated about three-dimensional Anti-de Sitter space is dual to a two-dimensional conformal field theory whose right- and left-moving central charges are given by $c_{R}=24k$ and $c_{L}=0$, respectively, being $k$ the level of the Chern-Simons action. We study the classical theory both at the linear and non-linear level. In particular, we show how Chiral Gravity is included as a special sector. In addition, the theory has other sectors, which we explore; we exhibit analytic exact solutions that are not solutions of Topologically Massive Gravity (and, consequently, neither of General Relativity) and still satisfy Brown-Henneaux asymptotically AdS$_{3}$ boundary conditions.
\end{abstract}

\maketitle

\section{Introduction}   
In Ref. \cite{Witten}, Witten proposed that three-dimensional quantum gravity on Anti-de Sitter (AdS) space could be dual to a two-dimensional conformal field theory (CFT) with holomorphic factorization. This proposal led to an elegant construction at the highly quantum level. However, soon after, the hypothesis of holomorphic factorization as holding for generic values of the coupling constants was criticized by some authors \cite{Gaberdiel, Gaiotto}. Still, it became clear that the idea of demanding the partition function to be holomorphically factorizable was an ingenious one, as such property would have been of help to overcome several problems encountered when trying to define a quantum version of general relativity (GR) in three dimensions \cite{MaloneyWitten}. In Ref. \cite{CG}, Li, Song and Strominger, reversing the approach, proposed a three-dimensional theory of gravity that, by construction, seemed to be dual to a holomorphic (chiral) CFT$_{2}$. The theory proposed in \cite{CG} is known as Chiral Gravity (CG), and it corresponds to Topologically Massive Gravity (TMG) \cite{TMG} formulated at a special point in parameter space, where the curvature radius of AdS$_{3}$, $l$, equals the inverse of the graviton mass, $\mu $. The consistency of this construction has been discussed in \cite{StromingerProof, GJ, CDWW, GKP, HMT, GAY, Maloney}. 

Although it became clear after Ref. \cite{Maloney} that, provided suitable boundary conditions are imposed, the CG theory may represent a consistent quantum gravity theory in AdS$_{3}$, some questions remain open, such as that about the relevant geometries that contribute to the partition function \cite{belgas}, or the question about how such a model could be embedded in string theory. Another interesting question is how to generalize CG\ in order to include more fields and local degrees of freedom, for instance. Here, we address the latter question. 

We propose a generalization of CG defined by a Chern-Simons action for a deformed Lorentz connection in 2+1 dimensions. The deformed connection consists of a torsion-free Riemannian connection $\tilde{\omega}$ and a conformal family of contorsion tensors $\phi\epsilon^a{}_{b \mu}$, where $\phi$ is a scalar field that describes the only\footnote{In TMG, there exists a local degree of freedom, corresponding to the massive graviton. However, at the chiral point and when Brown-Henneaux boundary conditions are imposed, the theory loses its local degree of freedom.} propagating degree of freedom of  the theory. The torsion-free condition for $\tilde{\omega}$ is enforced in a standard way through a Lagrange multiplier.

The paper is organized as follows: In Section II, the Lorentz-Chern-Simons theory is introduced, adding to it a constraint term for the torsion in such a way that the spin connection acquires a single additional mode in its contorsion part. Section III discusses how this model corresponds to a non-minimal extension of TMG at the chiral point, including CG as a particular sector. In Section IV, the field equations of the theory are obtained, including the constraint equation, both in the first-order and second-order formalisms. In Section V, the linear approximation of the theory is discussed, analyzing the linearized field equations around maximally symmetric backgrounds. Section VI discusses the theory at non-linear level, exhibiting exact solutions to the field equations, which present interesting geometrical features. We also discuss how the black hole solutions of GR are embedded in this model. In section VII, we speculate about the possibility that this theory be dual to a holomorphic CFT$_{2}$ reviewing, in particular, the computation of the AdS$_{3}$ black hole entropy from the viewpoint of the CFT$_{2}$. Section VIII contains our conclusions and further remarks.

\section{Lorentz-Chern-Simons theory}   
Let us consider the three-dimensional Chern-Simons (CS) Lagrangian 
\begin{equation}
\mathcal{L}_{CS}(\omega )=\omega^a{}_b \wedge d\omega^b{}_a + \frac{2}{3}\omega^a{}_b \wedge \omega^b{}_c \wedge \omega^c{}_a , \label{CSLagrangian}
\end{equation}
where $\omega^{ab}$ are the components of the spin connection 1-form $\omega^{ab}=\omega^{ab}{}_\mu dx^\mu$ on a three-dimensional manifold  $M_3$. Here, Latin characters correspond to Lorentz indices, while Greek characters are coordinate indices. We work in the Einstein-Cartan formalism where the spin connection $\omega^{ab}$ and the dreibein 1-form $e^a = e_\mu ^a dx^\mu$ are considered as independent dynamical fields on equal footing. The spin connection can be decomposed in two independent parts 
\begin{equation}
\omega ^{ab}=\tilde{\omega}^{ab}+\kappa ^{ab},  \label{SpinConnection}
\end{equation}
where the Riemannian (purely metric) connection $\tilde{\omega}^{ab}$ is defined to satisfy the torsion-free condition 
\begin{equation}
\tilde{D}e^{a}=de^{a}+\tilde{\omega}^{a}{}_{b}\wedge e^{b}\equiv 0,
\label{compatibility}
\end{equation}
{where $\tilde{D} =d + \tilde{\omega}$ denotes covariant derivative in the Riemannian connection,} and the contorsion tensor $\kappa^{ab}$ is related to the torsion 2-form, 
\begin{equation}
T^a=d e^a +\omega^a{}_b\wedge e^b = \kappa^a{}_b\wedge e^b . \label{T=De}
\end{equation}

{The field equations for the Lagrangian (\ref{CSLagrangian}) imply that all classical configurations are Lorentz-flat,
\begin{equation}
R^{ab}=0 . \label{LorentzFlat}
\end{equation}
As a direct consequence of this, the torsion 2-form must be covariantly constant, 
\begin{equation*}
DT^a= dT^a + \omega^a{}_b\wedge T^b = R^a{}_b\wedge e^b \equiv 0.
\end{equation*}
 In three dimensions, this equation can be integrated and the solution is
 \begin{equation}
 T^a = \phi_0 \epsilon^a{}_{bc}e^b\wedge e^c, \label{T=ee}
 \end{equation}
 where $\phi_0$ is a constant with dimension [length]$^{-1}$. Comparing (\ref{T=ee}) and (\ref{T=De}), the contorsion is found to be
\begin{equation}
\kappa^{ab}=-\phi_0\epsilon^{ab}{}_c e^{c}.\label{contorsion}
\end{equation}
Finally, combining (\ref{SpinConnection}), (\ref{contorsion}) and (\ref{LorentzFlat}) the Riemannian curvature $\tilde{R}^{ab} = d \tilde{\omega}^{ab} + \tilde{\omega}^a{}_c \wedge\tilde{\omega}^{cb}$ can be seen to be constant and negative,
\begin{equation}
\tilde{R}^{ab} = -(\phi_0)^2 e^a \wedge e^b. \label{R-tilde=-ee}
\end{equation}
In other words, all on-shell configurations obtained from (\ref{CSLagrangian}) are three-dimensional locally AdS spacetimes with constant torsion, where the cosmological constant $\Lambda=-(\phi_0)^2\leq 0 $ is an integration parameter  \cite{APRSZ}.}

We now consider a minimal deviation from the strictly covariantly constant torsion case. Presumably, this could be the result of some form of spinning matter that acts as a local source for torsion. One way to allow for this degree of freedom is by promoting $\phi_0$ to be a local dynamical field $\phi$, so that the connection now reads
\begin{equation}
\omega^{ab}=\tilde{\omega}^{ab} {-} \phi \epsilon^{ab}{}_c e^c. \label{SpinCon-phi}
\end{equation} 
Inserting (\ref{SpinConnection}) in the Lagrangian (\ref{CSLagrangian}) yields
\begin{align*}
\mathcal{L}_{CS}(\omega) &  =\mathcal{L}_{CS}(\tilde{\omega})+2\kappa^a{}_b \wedge\tilde{R}^b{}_a + \frac{2}{3}\kappa^a{}_b \wedge \kappa^b{}_c \wedge\kappa^c{}_a + \\
&  \kappa^a{}_b \wedge\tilde{D} \kappa^b{}_a + d \left(\tilde{\omega}^a{}_b \wedge\kappa^b{}_a \right)  ,
\end{align*}

The torsion-free CS Lagrangian can be expressed in terms of the Christoffel connection $\tilde{\Gamma}^\alpha_{\beta \mu} \equiv \{^\alpha_{\beta\mu}\}$,
\begin{equation*}
\mathcal{L}_{CS}(\tilde{\omega}) = \left[ \tilde{\Gamma}^\alpha_{\beta \mu} \partial_\nu \tilde{\Gamma}^\beta_{\alpha \rho} + \frac{2}{3}  \tilde{\Gamma}^\alpha_{\beta \mu} \tilde{\Gamma}^\beta_{\lambda \nu} \tilde{\Gamma}^\lambda_{\alpha \rho} \right] \epsilon^{\mu \nu \rho} d^3x. \label{L=L(Gamma)}
\end{equation*}

The choice (\ref{SpinCon-phi})\ reduces the number of independent components of $\kappa ^{ab}$ from nine to one. In this case the theory defined by Lagrangian (\ref{CSLagrangian}) resembles TMG non-minimally coupled to a scalar field, 
\begin{align}
I_{CS}& \equiv \frac{k}{4\pi }\int\limits_{M_{3}}\mathcal{L}_{CS}(\omega ) = \frac{k}{2\pi }\int\limits_{M_{3}}\bigg[\frac{1}{2}\mathcal{L}_{CS}(\tilde{\omega})+\phi ^{2}e_{a}\wedge \tilde{T}^{a}+  \notag \\
& \phi \epsilon _{abc}\left( \tilde{R}^{ab}+\frac{1}{3}\phi ^{2}e^{a}\wedge e^{b}\right) \wedge e^{c}+\frac{1}{2}d\left( \phi \epsilon _{abc}\tilde{\omega}^{ab}\wedge e^{c}\right) \bigg].  \label{S}
\end{align}
This expression could be further simplified by dropping the term involving $\tilde{T}^a=\tilde{D}e^a$ which vanishes by virtue of (\ref{compatibility}). However, we prefer to keep this term for future convenience, because in Section IV.B the constraint $\tilde{T}=0$ is implemented by means of a Lagrange multiplier, adding to (\ref{S}) the term
\begin{equation}
\frac{k}{8\pi }\int\limits_{M_{3}}\zeta _{a}\wedge \tilde{T}^{a},  \label{Lm}
\end{equation}
where $\zeta _{a}$ is a vector-valued 1-form. The variation with respect to $\zeta _{a}$ yields the compatibility condition (\ref{compatibility}).

In the expressions above, $k$ is the level of the Chern-Simons action, and is given by a positive integer number, 
\begin{equation*}
k\in \mathbb{Z}_{>0}.
\end{equation*}
The scalar field $\phi $ enters in (\ref{S}) as an effective cosmological (non-constant) term, and also acts as a non-constant Planck scale.

The theory defined by action (\ref{S}) {with} the constraint term (\ref{Lm}) represents a generalization of TMG \cite{TMG} (at the chiral point \cite{CG}). In fact, one can verify that any solution of the latter theory solves the equations of motion derived from (\ref{S}) for $\phi = const.$ In addition, as we will show, the theory contains more general dynamical sectors (with $d\phi \neq 0$) which exhibit new geometrical features; for instance, we will exhibit asymptotically AdS$_3$ solutions with non-constant curvature.

Before going further, let us consider a generalization of the action above by introducing two deformation parameters ($\lambda$ and $\mathfrak{m}$), which will allow to adjust the cosmological constant ($\Lambda$) and the mass parameter of the topologically massive term ($\mu$). Namely, we consider the action 
\begin{align}
I_{[\lambda ,\mathfrak{m}]}& =\frac{k}{2\pi }\int_{M_3}\bigg[\epsilon_{abc}\left(\phi \tilde{R}^{ab}\wedge e^c + \frac{\lambda}{3!} \phi^3 e^a \wedge e^b \wedge e^c \right) +  \notag \\
& \frac{1}{2\mathfrak{m}}\mathcal{L}_{CS}(\tilde{\omega})\bigg] + \frac{k}{2\pi } \int_{M_3}\bigg[\phi^2 e_a \wedge \tilde{T}^a + \frac{1}{2\mathfrak{m}} \zeta_a \wedge \tilde{T}^a +  \notag \\
& \frac{1}{2} d\left(\phi \epsilon_{abc}\tilde{\omega}^{ab}\wedge e^c \right) \bigg].  \label{I}
\end{align}

Theory (\ref{I}) clearly reduces to (\ref{S}) with (\ref{Lm}) for the special case $\lambda=2$, $\mathfrak{m}=1$; namely $I_{CS}=I_{[2,1]}$. 

\section{Chiral Gravity}     
Let us now see that the original case $\lambda=2\mathfrak{m}=2$, for a fixed value of $\phi$, corresponds to a generalization of TMG formulated at the so-called chiral point of the parameter space \cite{CG}, in the sense that {for $\phi=\phi_0$ the two Lagrangians are the same. In order to see} this, let us first assume that $\phi$ takes the value $\phi_0$ in certain limit (say close to the boundary in asymptotically Anti-de Sitter space). Then, since $\lambda =2$, the first two terms in (\ref{I}) become\footnote{Our conventions are such that $g_{\mu\nu} =\eta_{ab} e^a_{\mu} e^b_\nu$ with $\eta_{ab}=diag(-,+,+)$. The inverse relation is $g^{\mu\nu} = \eta^{ab}E_a^\mu E_b^\nu$, where $e^a{}_\mu E_a{}^\nu = \delta_\mu ^\nu$. The (Riemannian) curvature two-form is given by $\tilde{R}^{ab}=(1/2)\tilde{R}^{ab}{}_{\mu \nu} dx^\mu \wedge dx^\nu$.} 
\begin{equation*}
\frac{k}{2\pi }\int_{M_3}\epsilon_{abc}\phi \tilde{R}^{ab}\wedge e^c = - \frac{k\phi_0}{2\pi}\int_{M_3}d^3x \sqrt{-g}\tilde{R}
\end{equation*}
and 
\begin{equation*}
\frac{k}{6\pi }\int_{M_3}\epsilon_{abc}\phi^3 e^a \wedge e^b \wedge e^c = -\frac{k\phi_0 ^3}{\pi}\int_{M_3} d^3 x\sqrt{-g}
\end{equation*}
\newline
respectively. Comparing these formulas with the standard expressions
\begin{equation*}
\frac{-1}{16\pi G}\int_{M_3}d^3 x\sqrt{-g}\tilde{R}\quad \mbox{and} \quad \frac{\Lambda}{8\pi G}\int_{M_3}d^3 x\sqrt{-g},
\end{equation*}
allows identifying the effective three-dimensional Newton constant as $G = 1/(8k\phi_0)$, and the effective cosmological constant as $\Lambda \equiv -l^{-2}=-8 G k\phi_0^3 $, and therefore $l^{2}=1/\phi_0 ^2$, in agreement with (\ref{R-tilde=-ee}).

On the other hand, comparing with the standard topologically massive term, 
\begin{equation*}
\frac{1}{32\pi G\mu} \int_{M_3} d^3 x\epsilon^{\mu \nu \rho}\tilde{\Gamma}_{\mu \alpha}^\eta \left(\partial_\nu \tilde{\Gamma}_{\rho \eta} ^\alpha + \frac{2}{3} \tilde{\Gamma}_{\nu \beta}^\alpha \tilde{\Gamma}_{\rho \eta }^\beta \right) 
\end{equation*}%
implies $k/4\pi =1/(32\pi G\mu )$. Combining these identifications, the following relations are found
\begin{eqnarray}
\mu  &=&\phi_0 =\pm \frac{1}{l},  \label{chiralpoint} \\
k &=&\pm \frac{l}{8G}  \label{k}.
\end{eqnarray}
Equation (\ref{chiralpoint}) defines the so-called chiral point of TMG, and, for this choice of couplings, the theory formulated about asymptotically AdS$_3$ is referred to as Chiral Gravity. TMG at (\ref{chiralpoint}) exhibits special features and it was proposed as a candidate for a consistent quantum theory \cite{CG}.

\section{Field equations}        
\subsection{The Chern-Simons theory}   
Including the term  (\ref{Lm}) in the action (\ref{S}) breaks conformal symmetry; without this term it {would be} possible to absorb $\phi$ in a redefinition of the dreibein,
\begin{equation}
\theta^a \equiv \phi e^a , \label{weyl rescaling}
\end{equation}
{eliminating the  scalar field from the Lagrangian. Then}, in terms of the new dreibein $\theta^a$, the field equations obtained varying with respect to $\theta ^{a}$ and $\tilde{\omega}^{ab}$ are 
\begin{align*}
\epsilon_{abc}\tilde{R}^{ab} +\frac{1}{2}\lambda \epsilon_{abc}\theta^a \wedge \theta^b + 2\tilde{\tau}_c & =0,\\ 
-\frac{1}{\mathfrak{m}}\tilde{R}^{ab} + \epsilon^{ab}{}_c \tilde{\tau}^c -\theta^a \wedge \theta^b & = 0.  
\end{align*}
Here the 2-form $\tilde{\tau}={\frac{1}{2}}\tilde{\tau}_{\mu \nu}dx^\mu \wedge dx^\nu$ is the torsion defined with the rescaled basis (\ref{weyl rescaling}) and $\tilde{\omega}$, namely $\tilde{\tau}^a =d\theta^a + \tilde{\omega}^a{}_b \wedge \theta^b$. These equations can also be written as 
\begin{align}
\epsilon_{abc}\left( \tilde{R}^{ab} + \frac{\lambda}{2}\theta^a \wedge \theta^b \right) +2\tilde{\tau}_c & =0,  \label{eom without const 4} \\
\epsilon_{abc}\left(\frac{1}{\mathfrak{m}}\tilde{R}^{ab}+ \theta^a \wedge \theta^b \right) + 2\tilde{\tau}_c & =0,  \label{gggg}
\end{align}
which, for $\lambda \neq 2\mathfrak{m}$, can be solved for $\tilde{R}$ and $\tilde{\tau}$, 
\begin{align*}
\tilde{R}^{ab}& =\left( \frac{\mathfrak{m}-\lambda /2}{\mathfrak{m}-1} \right) \,\theta ^{a}\wedge \theta ^{b}, \\
\tilde{\tau}_c & =-\frac{1}{2}\left( \frac{\mathfrak{m}-\lambda /2}{\mathfrak{m}-1}\right) \epsilon_{abc}\, \theta^a \wedge \theta^b.
\end{align*}

This means that the solutions of this system have locally constant curvature and constant torsion. This system corresponds to the Mielke-Baekler theory \cite{MB}. In the case $\mathfrak{m}=\lambda /2$ and $\tilde{\tau}_{c}=0$, equations (\ref{eom without const 4}) and (\ref{gggg}) coincide and the theory degenerates. On the other hand, the even more special case $\mathfrak{m}=\lambda /2=1$ is similar to the one studied in Ref. \cite{Torsed}.

\subsection{Implementing the constraint}     
The field equations for the theory defined by (\ref{S}) with the addition of the constraint term (\ref{Lm}), are
\begin{align}
0 = &\phi \epsilon_{abc}\tilde{R}^{ab}+\frac{\lambda}{2} \epsilon_{abc}\phi^3 e^a \wedge e^b +2\phi d\phi \wedge e_c +  \notag \\
& \; 2\phi^2 \tilde{T}_c +\frac{1}{2\mathfrak{m}}\tilde{D}\zeta_c, \label{19} \\
0 =&-\frac{1}{\mathfrak{m}}\tilde{R}^{ab} + \epsilon^{ab}{}_c \tilde{D}\left( \phi e^c \right) -\phi^2 e^a \wedge e^b - \frac{1}{2\mathfrak{m}} \zeta^{[ a} \wedge e^{b]},  \label{20} \\
0  =&\epsilon _{abc}\tilde{R}^{ab}\wedge e^c +\frac{\lambda}{2}\phi^2 \epsilon_{abc} e^a \wedge e^b \wedge e^ c +2\phi e_a \wedge \tilde{T}^a ,  \label{21} \\
0 = &\frac{1}{2\mathfrak{m}}\tilde{T}^a ,  \label{22}
\end{align}
obtained by varying with respect to $e^a $, $\tilde{\omega}^{ab}$, $\phi$ and $\zeta_a$, respectively. Bracketed indices denote normalized
antisymmetrization $A_{[ab]}=\frac{1}{2}(A_{ab}-A_{ba})$ {and (\ref{22}) is} the compatibility condition (\ref{compatibility}).

Eq. (\ref{20}) can be algebraically solved for $\zeta^{a}$ by applying systematically the contraction operator $\iota_a$, defined to act on a $p$-form as $\iota_a p=\frac{1}{p!}E_{a}^{\ \mu}p_{\mu\mu_1\cdots\mu_{p-1}}dx^{\mu_1}\wedge\cdots\wedge dx^{\mu_{p-1}}$. We obtain
\begin{equation*}
\zeta ^{a}=4\mathfrak{m}\left( -B^{a}+\frac{1}{4}Be^{a}\right) \,,
\end{equation*}
{where we have defined}
\begin{align*}
B^a & = -\frac{1}{\mathfrak{m}}\tilde{R}^a + \epsilon^{ab}{}_c \partial_b \phi e^c + 2\phi^2 e^a \\
B& = -\frac{1}{\mathfrak{m}}\tilde{R}+ 6\phi^2,
\end{align*}
{and}
\begin{align*}
\tilde{R}^{a}=\iota_{b}\tilde{R}^{ab}=E_{b}^{\ \mu}\tilde{R}^{ab}{}_{\nu\mu}dx^\nu \\
\tilde{R}=\iota _{a}\tilde{R}^{a} = E_{a}^{\ \nu}E_{b}^{\ \mu}\tilde{R}^{ab}{}_{\nu\mu}.
\end{align*}

Substituting {$\zeta$ in (\ref{19}) and solving (\ref{22}) for $\tilde{\omega}$}, gives a system of third order differential equations for $e^a_{\ \mu}$
\begin{equation}
\frac{1}{2}\phi \epsilon_{abc}\left(\tilde{R}^{ab} +\frac{\lambda}{2}\phi^2 e^a \wedge e^b \right) +\frac{1}{\mathfrak{m}} C_c -\tilde{D} \ast \left(d\phi \wedge e_c \right) =0, \label{equation multiplier free}
\end{equation}
where we have defined the Cotton 2-form
\begin{equation}
C^a \equiv \tilde{D}\left( \tilde{R}^a -\frac{1}{4}\tilde{R}e^a \right) ,  \label{Cotton two form}
\end{equation}
and $\ast $ stands for the Hodge dual.\footnote{Our convention is such that $\ast \left(e^{a_1}\wedge \cdots \wedge e^{a_p} \right) = \frac{1}{\left( D-p\right) !}\epsilon^{a_1 \cdots a_p}{}_{a_{p+1}\cdots a_D} e^{a_{p+1}}\wedge \cdots \wedge e^{a_D}$.} The system is now given by (\ref{21}) and (\ref{equation multiplier free}), and remains to be solved for the dreibein and the scalar field.

Contracting {Eq. (\ref{equation multiplier free}) with $e^c$ and using the identity $C^{a}\wedge e_a =0$ combined with (\ref{21}), one finds that the scalar field is classically a harmonic function,} 
\begin{equation*}
d\ast d\phi=0,
\end{equation*}
(cf. Eq. (\ref{eom4}) below).

\subsection{Metric formulation}    
The theory can be conveniently studied in the second-order formalism, where the fields are the metric $g_{\mu\nu}=e^a_\mu e^b_\nu \eta_{ab}$ and the scalar field $\phi$. {In this case, the equations obtained from the reduced action, where the torsion has been set to zero, are equivalent to those obtained in the first order form (\ref{19}-\ref{22}). As shown above, equations (\ref{20}) and (\ref{22}), obtained by varying the original first-order action with respect to $\tilde{\omega}$ and $\zeta$ respectively, can be algebraically solved for these auxiliary fields, $\tilde{\omega}=\tilde{\omega}(e^a, \phi)$ and $\zeta=\zeta(e^a, \phi)$. Then, the reduced action in which these expressions for $\tilde{\omega}$ and $\zeta$ have been used, yields the same equations for $e^a$ and $\phi$ (see, e.g. \cite{Henneaux-Teitelboim}).}

{Let us consider the Hodge} dual of Eq. (\ref{equation multiplier free}),
\begin{align}
0 & =\phi\left( \tilde{R}_{\mu\nu}-\frac{1}{2}g_{\mu\nu}\tilde{R}\right) -\frac{\lambda}{2}\phi^3 g_{\mu\nu} + \frac{1}{\mathfrak{m}} \tilde{C}_{\mu\nu} +  \notag \\
& g_{\mu\nu}\tilde{\nabla}_\alpha\tilde{\nabla}^\alpha \phi - \tilde{\nabla}_\mu \tilde{\nabla}_\nu \phi\, , \label{eom1}
\end{align}
where, symbolically, $\tilde{\nabla}=\partial+\tilde{\Gamma}$ is the covariant derivative for the Christoffel connection and $\tilde{C}_{\mu\nu}$ is now the Cotton tensor, defined by
\begin{equation*}
\tilde{C}^\mu{}_\nu =\frac{1}{2}\epsilon^{\mu\alpha\beta}\tilde{\nabla}_{\alpha}\tilde{R}_{\beta\nu} +\frac{1}{2} \epsilon_\nu{}^{\alpha\beta} \tilde{\nabla}_\alpha \tilde{R}^\mu{}_\beta,   \label{Cotton tensor}
\end{equation*}
that can also be written as a derivative of the Schouten tensor, 
\begin{equation*}
\tilde{C}_{\mu\nu}=\epsilon_\mu{}^{\alpha\beta}\tilde{\nabla}_\alpha \left(\tilde{R}_{\beta\nu}-\frac{1}{4}g_{\beta\nu}\tilde{R}\right) ,
\end{equation*}
cf. Eq. (\ref{Cotton two form}). On the other hand, (\ref{21}) reads
\begin{equation}
\tilde{R}+3\lambda\phi^{2}=0.  \label{eom2}
\end{equation}
By taking the trace of (\ref{eom1}) and considering (\ref{eom2}), one finds the equivalent set of equations 
\begin{equation}
\phi\left(\tilde{R}_{\mu\nu}-\frac{1}{2}g_{\mu\nu}\tilde{R}\right) -\frac{\lambda}{2} \phi^3 g_{\mu\nu}+\frac{1}{\mathfrak{m}} \tilde{C}_{\mu\nu} - \tilde{\nabla}_\mu \tilde{\nabla}_\nu \phi=0,
\label{eom3}
\end{equation}
and the harmonic equation 
\begin{equation}
\tilde{\nabla}_\mu \tilde{\nabla}^\mu \phi=0.  \label{eom4}
\end{equation}
From this, it follows that in the case $\phi=const$ equations (\ref{eom3})-(\ref{eom4}) reduce to Topologically Massive Gravity \cite{TMG} with cosmological constant given by $\Lambda=-\lambda \phi_0^2 /2$ and graviton mass $\mu = \mathfrak{m}\phi_0$. 

\section{Linearized theory}  
Let us now study the linearized theory as a perturbation of the metric and the scalar field about a given solution $\bar{g}_{\mu \nu }$, $\bar{\phi}$ of (\ref{eom1}, \ref{eom2}). We consider
\begin{align}
g_{\mu \nu }& =\bar{g}_{\mu \nu }+h_{\mu \nu }, \notag \\
\phi & =\bar{\phi}+\varphi. \notag
\end{align}

The first order corrections of Eqs. (\ref{eom1}, \ref{eom2}) are
\begin{align}
\varphi \bar{G}_{\mu \nu }+\bar{\phi}G_{\mu \nu }^{(1)}-\frac{1}{2}\lambda \bar{\phi}^3 h_{\mu \nu }-\frac{3}{2}\lambda \bar{\phi}^2 \varphi \bar{g}_{\mu \nu }&   \notag \\
+ \frac{1}{\mathfrak{m}}C_{\mu \nu }^{(1)}-\gamma_{\mu \nu }^\lambda \partial_\lambda \bar{\phi}-\bar{\nabla}_\mu \bar{\nabla}_\nu
\varphi & =0,  \label{gggggg}\\
\bar{\nabla}_\mu \bar{\nabla}_\nu h^{\mu\nu}-\bar{\nabla}^2 h -h^{\mu\nu}\bar{R}_{\mu\nu}+6\lambda \bar{\phi}\varphi&=0 \label{ggggg}
\end{align}
where $\gamma^\lambda_{\mu \nu}$ stands for the first-order correction to the Christoffel symbol, $\tilde{\Gamma}^\lambda_{\mu \nu} =\bar{\Gamma}^\lambda_{\mu\nu} +\gamma^\lambda_{\mu \nu} +\mathcal{O}( h^2 ) $, with 
\begin{equation*}
\gamma^\lambda _{\mu \nu} =\frac{1}{2}\left(\bar{\nabla}_\mu h^\lambda{}_\nu+ \bar{\nabla}_\nu h^\lambda{}_\mu -\bar{\nabla}^\lambda h_{\mu\nu}\right) ,
\end{equation*}
{where indices are raised and lowered with the background metric $\bar{g}^{\mu\nu}$, $\bar{g}_{\mu\nu}$.}

The first-order corrections for the Ricci tensor $\tilde{R}_{\mu\nu}=\bar{R}_{\mu\nu}+R_{\mu\nu}^{(1)}$, and Ricci scalar $\tilde{R}=\bar{R}+R^{(1)}$, are given by 
\begin{align*}
R_{\mu\nu}^{(1)} & =\frac{1}{2}\left( 2\bar{\nabla}_\lambda \bar{\nabla }_{(\mu}h^{\lambda}{}_{\nu)} -\bar{\nabla}_\mu \bar{\nabla}_\nu h-\bar{\nabla}^2 h_{\mu\nu}\right) ,\\
R^{(1)} & =\bar{\nabla}_\mu \bar{\nabla}_\nu h^{\mu\nu} -\bar{\nabla}^2 h -h^{\mu\nu}\bar{R}_{\mu\nu},
\end{align*}
from which one can build the first-order corrections of the Einstein and Cotton tensors, $\tilde{G}_{\mu\nu}=\bar{G}_{\mu\nu}+G^{(1)}_{\mu\nu}$, $\tilde{C}_{\mu\nu}=\bar{C}_{\mu\nu}+C^{(1)}_{\mu\nu}$,
\begin{align*}
G_{\mu\nu}^{(1)} =\; & R^{(1)}_{\mu\nu}-\frac{1}{2}(\bar{g}_{\mu\nu}R^{(1)}+h_{\mu\nu}\bar{R}),& \notag \\
C_{\mu\nu}^{(1)} =\; & \epsilon_{(\mu|}{}^{\alpha\beta}\left( \bar{\nabla}_\alpha R^{(1)}_{\beta|\nu)}+\gamma^\lambda_{\nu)\alpha}\bar{R}_{\lambda\beta} \right)&\notag\\
& + 2\epsilon_{(\mu|\lambda}{}^{[\alpha}h^{\beta]\lambda}\bar{\nabla}_\alpha\bar{R}_{\beta|\nu)}.&
\end{align*}
Here parenthesis denote normalized symmetrization $A_{(\mu\nu)}=\frac{1}{2}(A_{\mu\nu}+A_{\nu\mu}).$

\subsection{Gauge fixing}   
{In order to identify the physical degrees of freedom it is necessary to separate the gauge degrees of freedom from the propagating components of the fields. This is usually achieved by making a gauge transformation in which the new metric is transverse ($\bar{\nabla}_\mu h'^{\mu\nu}=0$) and traceless ($h'=0$). In the metric form of the theory, the local Lorentz symmetry is gone since the fields $g_{\mu \nu}$ and $\phi$ are trivially Lorentz invariant and the only remaining gauge invariance is diffeomorphism symmetry. Under an infinitessimal  diffeomorphism parametrized by $\xi^\mu$, the metric transforms as {$g_{\mu \nu}\rightarrow g_{\mu \nu} + \nabla_{(\mu}\xi_{\nu)}$.} In the linearized approximation, this corresponds to a change in $h_{\mu \nu}$ and $\varphi$ given by\footnote{{Note that in order for this transformation to be compatible with the linearized approximation, $\xi_\mu$ must be of the same order as $h_{\mu \nu}$}.}
\begin{equation*}
\delta_{\xi}h_{\mu\nu}=\bar{\nabla}_{(\mu}\xi_{\nu)}, \quad \delta_\xi \varphi = \xi^\mu \partial_\mu \varphi. \label{isometry}
\end{equation*} 

If the transformed field is transverse and traceless, the diffeomorphism $\xi $ must be such that}
 \begin{align}
\bar{\nabla}_\mu h'^{\mu\nu} = \bar{\nabla}_\mu h^{\mu\nu}+\bar{\nabla}_\mu \bar{\nabla}^{(\mu}\xi^{\nu)} &  =0,\label{transv gauge} \\
h' = h + \bar{\nabla}_{\mu}\xi^{\mu} &  =0.\label{hjk}
\end{align}

{In order for the transverse-traceless gauge to be accessible, these equations for $\xi$ must be integrable. Using the commutation relation of the covariant derivatives together with (\ref{hjk}), equation (\ref{transv gauge}) can be written as
\begin{equation*}
\frac{1}{2}\left(  \bar{\nabla}^{2}\xi^{\nu}+\bar{R}_{\mu}^{\ \ \nu}\xi^\mu \right)  +\bar{\nabla}_{\mu}h^{\mu\nu}-\frac{1}{2}\bar{\nabla}^{\nu}h= 0.\label{divergenless gauge 1}
\end{equation*}

Taking the divergence of this expression one finds $\bar{\nabla}_{\mu }\bar{\nabla}_{\nu }h^{\mu \nu }-\bar{\nabla}^{2}h+\bar{\nabla}_\mu\left(\bar{R}^\mu{}_\nu\xi^\nu\right)=0$. For an AdS background, in particular, it reads 
\begin{equation}
\bar{\nabla}_\mu\bar{\nabla}_\nu h^{\mu\nu}-\bar{\nabla}^2h+\frac{2}{l^2}h=0, \label{X}
\end{equation} 
which is incompatible with (\ref{ggggg}) if $\varphi\neq0$. This means that the transverse-traceless condition cannot be met in general, starting from a generic $h_{\mu \nu}$ and $\varphi$, because (\ref{transv gauge}) and (\ref{hjk}) are not integrable unless $\phi=const$.}

{On the other hand, a purely transverse gauge condition {$\bar{\nabla}_\mu h^{\prime\mu\nu}=0$} is allowed by (\ref{ggggg}) provided
\begin{equation}
\Big(\bar{\nabla}^2-\frac{2}{l^2}\Big)h=6\lambda\bar{\phi}\varphi, \label{Y}
\end{equation}
which is not contradictory if {$h^{\prime\mu \nu}$} is not traceless Then, we find that the trace of the perturbation $h=\bar{g}^{\mu\nu}h_{\mu\nu}$ is sourced by the perturbation of the scalar field $\varphi$ which is in turn a harmonic function
\begin{equation*}
\bar{\nabla}_\mu \bar{\nabla}^\mu \varphi = 0.
\end{equation*}

In conclusion, the presence of the scalar field excites a new degree of freedom associated to $h$.

\section{The non-linear theory}         
\subsection{Gravitational waves}   
Now, let us study non-linear solutions to the equations of motion.

In the sector $\phi=const$, theory (\ref{eom3})-(\ref{eom4}) reduces to Topologically Massive Gravity with a cosmological constant $\Lambda =-l^{-2}=-\lambda\phi_{0}^{2}/2\leq0$. In particular, it admits as an exact solution three-dimensional Anti-de Sitter space whose metric, in Poincar\'{e} coordinates, reads
\begin{equation}
ds^{2}=\frac{l^{2}}{y^{2}}\left( -du^{2}-2du\,dv+dy^{2}\right) . \label{AdS}
\end{equation}

A particularly interesting deformation of Anti-de Sitter solution (\ref{AdS}) is given by the so-called AdS-waves, which correspond to a particular case of the family of Siklos solutions of Einstein equations. {The AdS-wave \textit{ansatz}, is
\begin{equation}
ds^{2}=\frac{l^{2}}{y^{2}}\left( -F(u,y)du^{2}-2du\,dv+dy^{2}\right) , \label{AdS Waves Ansatz}
\end{equation}
which represents a $pp$-wave propagating on AdS$_{3}$ space where $F(u,y)$ describes the profile of the wave. 

The equations for $F(u,y)$ demand the scalar field to be constant. This is because the Ricci scalar for (\ref{AdS Waves Ansatz}) is $R=-6/l^{2}$ (c.f. (\ref{eom2})).} Then, all solutions (\ref{AdS Waves Ansatz}) reduce to the one studied in Ref. \cite{Eloy} and no deformation of this type gives rise dynamics for $\phi (x)$. In the next section, we consider solutions of non-constant $\phi $, which do not reduce to the TMG\ solutions.

\subsection{Circularly symmetric solutions} 
Now, let us consider circularly symmetric static solutions. Consider the diagonal form 
\begin{equation*}
ds^{2}=-f^{2}(r)dt^{2}+h^{2}(r)dr^{2}+r^{2}d\theta^{2}
\end{equation*}
where $r\in\mathbb{R}_{\geq0}$, $t\in\mathbb{R}$, and $\theta\in\lbrack 0,2\pi)$.

The system of differential equations for the radial metric functions $f$, $h$ and $\phi$ read 
\begin{align}
0 & =\frac{\phi h'}{rh^{3}}+\frac{1}{2}\lambda\phi^{3}+\frac{f' \phi' }{fh^{2}},  \label{eom cotton} \\
0 & =\frac{\phi f'}{rfh^2 }-\frac{1}{2}\lambda\phi^3 - \frac{1}{h}\left( \frac{\phi'}{h}\right)^{\prime}, \\
0 & =\frac{\phi}{hf}\left( \frac{f'}{h}\right)^{\prime} - \frac {1}{2}\lambda\phi^3 -\frac{\phi'}{rh^2 }, \\
0 & =r^2 \left[ \frac{1}{fh}\left( \frac{f'}{h}\right)^{\prime} + \frac{h'}{rh^3}\right]^{\prime}+ r\left[\frac{1}{fh}\left( \frac{f'}{h}\right) ^{\prime} - \frac{f' }{rfh^2}\right]  \notag \\
& +\frac{r^2 f'}{f}\left(\frac{h'}{rh^3} + \frac{f' }{rfh^2}\right) ,   \label{ggg}
\end{align}
where the primes stand for derivatives with respect to the radial coordinate, $f' \equiv\partial_r f$, etc.

The harmonic condition of the scalar field, on the other hand, takes the form 
\begin{equation}
\left( \frac{rf}{h}\phi ^{\prime }\right) ^{\prime }=0.  \label{harmonic}
\end{equation}

The system (\ref{eom cotton})-(\ref{harmonic}) admits an exact solution of the form 
\begin{equation}
ds^{2}=-\frac{r^2}{l^2}dt^2 + \frac{2r}{\phi_0 ^2 \lambda (r^2 -r_0^2 )^{3/2}}dr^2 +r^2 d\theta^2, \label{metric solution circular}
\end{equation}
with $r \geq r_0$, $t\in {\mathbb{R}}$ and $\theta \in [0,2\pi )$, and with a scalar field configuration
\begin{equation}
\phi (r)=\frac{\phi_0 (r^2 - r_0 ^2)^{1/4}}{r^{1/2}}, \label{phi solution circular}
\end{equation}
where $l$, $\phi _{0}$ and $r_{0}$ are integration constants (notice that, however, by rescaling $t$, we can set $l^2 \equiv 2/(\phi_0 ^2 \lambda)$ without loss of generality). {The asymptotic value of the scalar field at infinity is}
\begin{equation}
\lim_{r\rightarrow \infty}\phi (r)=\phi_0 .  \label{62}
\end{equation}

{The metric (\ref{metric solution circular}) is not defined if $\phi=\phi_0=0$. For $r_0 \rightarrow 0$ the scalar field approaches $\phi=\phi_0\neq 0$ and metric (\ref{metric solution circular}) approaches a locally AdS$_3$ geometry that corresponds to the massless BTZ solution \cite{BTZ, BHTZ}. In other words, the scalar field is not an independent hair\footnote{Nevertheless, it still represents a one-parameter family of static circularly symmetric solutions; the real parameter being $r_0$.} as it cannot be switched off; and when it approaches the constant value $\phi_0$, the generic solution becomes a particular black hole solution of TMG. Something similar occurs in other cases where scalar fields are supported by a black hole, in which case it is impossible to switch off the scalar keeping the mass of the black hole fixed \cite{HMTZ, MTZ}.} 

For large $r$ the metric (\ref{metric solution circular}) becomes
\begin{equation}
ds^2 \simeq -\frac{r^2}{l^2}dt^2 +\frac{2}{\phi_0 ^2 \lambda r^2 } dr^2 + r^2 d\theta^2 +...  \label{63}
\end{equation}
where the ellipses stand for terms that are subleading in powers of $r$. In fact, solution (\ref{metric solution circular})-(\ref{phi solution circular}) is asymptotically Anti-de Sitter space satisfying the Brown-Henneaux conditions \cite{BH}, which in particular require
\begin{align}
g_{tt}& \simeq -\frac{r^2 }{l^2 } + \mathcal{O}\left(r^0 \right) ,\qquad g_{rr}\simeq \frac{l^2}{r^2} +\mathcal{O}\left( r^{-4}\right) ,\label{Brown} \\
g_{\theta t}& \simeq \mathcal{O}\left( r^0 \right) ,\qquad g_{\theta \theta}\simeq r^2 +\mathcal{O}\left(r^0 \right) .  \label{Henneaux}
\end{align}

The theory actually admits other sets of boundary conditions, including logarithmic fall-offs $\sim \log (r)$ in the components above. This is known to happen in the CG\ theory \cite{GJ, GAY}, where imposing such behavior leads to the definition of the so-called Log-Gravity \cite{Maloney}. An important ingredient in the discussion in \cite{GJ, GKP, GAY,Maloney} was whether it is consistent to define the TMG\ theory at the chiral point imposing the strong conditions (\ref{Brown})-(\ref{Henneaux}). In \cite{Maloney, belgas}, the question whether non-constant curvature solutions obeying (\ref{Brown})-(\ref{Henneaux}) actually existed was studied in relation to the contributions of the Chiral Gravity partition function. This is why the fact of having found here non-constant curvature solutions obeying such strong fall-off behavior is relevant. 

{In principle, metric (\ref{phi solution circular}) can be considered also in the region $r<r_0$. However, the geometry turns out to be singular at $r=r_0$. This can be seen by computing the components of the Riemann tensor, which in three dimensions is given in terms of the metric and the Ricci tensor. At $r= r_0$, both the metric and the Ricci tensor exhibit singularities; in particular, $R_{rr}=-(2r^2+r_0^2)/(r^2(r^2-r_0^2))$. In addition, spacetime (\ref{metric solution circular}) also presents a singularity at $r=0$. Provided $r_0 \neq 0$, the scalar curvature invariants associated diverge for $r \to 0$ as $R\sim 1/r$, $R_{\mu\nu}R^{\mu\nu}\sim 1/r^6$, $R^\mu{}_\alpha R^\alpha{}_\beta R^\beta{}_\mu\sim 1/r^9$ and $\nabla_\mu R^{\alpha\beta}\nabla_\alpha R^\mu{}_\beta\sim1/r^9$, while all of them vanish at $r=r_{0}$.}


\subsection{Black holes}
Let us now consider black hole solutions. Since the theory includes TMG\ as a particular sector, it also exhibits black holes; in particular, the BTZ black hole \cite{BTZ}.

In the sector $\phi =\phi _{0}=const.$, equations (\ref{eom cotton})-(\ref{ggg}) simplify considerably and can be shown to admit solutions with
\begin{equation*}
f(r)=h^{-1}(r)=\left( \frac{r^2 -r_{+}^2}{l^2}\right)^{1/2},\qquad l^2 =\frac{2}{\lambda \phi_0 ^2}
\end{equation*}
where $r_{+}^{2}$ is a real constant. This corresponds to the BTZ black hole \cite{BTZ, BHTZ}
\begin{equation*}
ds^2 =-\frac{r^2 - r_{+}^2}{l^2} dt^2 + \frac{l^2}{r^2 - r_{+}^2} dr^2 + r^2 d\theta^2 ,
\end{equation*}
with $\phi =\phi _0$. The integration constant $r_{+}$ represents the location of the black hole horizon. The metric of the black hole solution that includes rotation reads
\begin{align*}
ds^2 = &-\frac{(r^2 -r_{+}^2 )(r^2 -r_{-}^2 )}{l^2 r^2}dt^2 +\frac{l^2 r^2dr^2}{(r^2 -r_{+}^2)(r^2 -r_{-}^2)}  \notag \\
&+ r^2 \left(d\theta^2 + \frac{r_{+}r_{-}}{lr^2}dt \right)^2,
\end{align*}
where $r_{-}$ represents the location of the inner horizon.

For this geometry, one can define two temperature parameters
\begin{equation}
T_{\pm}=\frac{1}{2\pi l^{2}}(r_{+}\pm r_{-}),   \label{m}
\end{equation}
which are associated to the inverse of the identification periods of the orbifold construction \cite{BHTZ}. In particular, this gives the Hawking
temperature
\begin{equation*}
T_{\text{H}}=T_{+}+T_{-}.   
\end{equation*}

In TMG, the expression of the black hole entropy does not satisfy the Bekenstein-Hawking area law, but it involves as well the area of the inner horizon; namely, one finds that the entropy is given by
\begin{equation}
S_{\text{BH}}=\frac{2\pi(r_{+}-r_{-})}{4G}.   \label{SBH}
\end{equation}
In the next section we will review how this result can be obtained from a dual CFT$_2$ point of view.

\section{AdS$_{3}$/CFT$_{2}$}    

In the theory we have defined here, the effective cosmological constant $l^{-2}$ (i.e. the inverse of the curvature radius of its AdS$_{3}$ solutions) enters as an integration constant, associated to the boundary value $\phi _{0}$ (see, for instance, Eqs. (\ref{62}) and (\ref{63})). A priori, this could seem surprising and, when thought of within the context of AdS$_{3}/$CFT$_{2}$ correspondence \cite{Malda}, it could even seem puzzling. This is because the central charge of the dual conformal field theory is typically given in terms of the curvature radius $l$ \cite{BH}. Then, if $l$ is free to take an arbitrary value within a continuous range, this may seem to contradict the Zamolodchikov $c$-theorem \cite{Zamolodchikov}, which forbids the existence of a family of CFT$_2$s parametrized by continuous values of the central charge. However, this is not a problem here because, although $l$ may take values on a continuum, the ratio $l/G$, which is what actually enters in the central charge, only takes specific (non-continuous) values, see (\ref{k}). That is, the theory happens to circumvent the obstruction imposed by the $c$-theorem and still present an infinite family of AdS$_{3}$ vacua parameterized by a continuous parameter $l$. Then, we conjecture that the theory in AdS$_{3}$ is holographically dual to a CFT$_{2}$ with left- and right-moving central charges given by
\begin{equation}
c_{L}=0\text{ \qquad}c_{R}=\frac{3l}{G}=24k.   \label{c}
\end{equation}

Indeed, this can be seen to be the case for a theory defined with $\phi = const.$ \cite{GonzalezSalgado}.

Modular invariance of such a CFT$_{2}$ would require%
\begin{equation*}
c_{R}-c_{L}=24k\in\mathbb{Z}_{\geq0},
\end{equation*}
which is actually satisfied due to the quantization of the CS\ level.

The temperature parameters (\ref{m}) read $T_{\pm }=(\phi _{0}^{2}/2\pi)(r_{+}\pm r_{-})$, and the black hole entropy (\ref{SBH})\ is $S_{\text{BH}}=4\pi k\phi _{0}(r_{+}-r_{-})$. Then, one comes to the conclusion that, for (\ref{c}), Cardy formula in the dual CFT$_{2}$ reads \cite{Cardy}
\begin{equation*}
S_{\text{CFT}}=\frac{\pi ^{2}l}{3}\left( c_{L}T_{+}+c_{R}T_{-}\right) , 
\end{equation*}
exactly reproducing the black hole entropy (\ref{SBH}).

\section{Conclusions and further remarks}    
{Here, a generalization of the theory of Chiral Gravity has been proposed. The model follows from considering a Chern-Simons action for the spin connection, supplemented with a scalar field that plays the role of a cosmological ``constant", and a constraint that enforces the spin connection to remain torsionless. This introduces a local degree of freedom in the theory, which incarnates as a scalar field non-minimally coupled to the metric. The theory includes TMG and Chiral Gravity of \cite{CG} as particular sectors}.

In this theory, effective cosmological constant, --{i.e.} the curvature radius of the maximally symmetric solutions--, appears as an integration constant related to the value of the contorsion at infinity. Its value is either negative or zero. In the former case, the theory admits an infinite family of Anti-de Sitter (AdS) vacua, labeled by a continuous parameter. We explained how this fact is not in conflict with Zamolodchikov's $c$-theorem in the dual conformal field theory (CFT). In fact, we conjecture that the theory on its AdS$_3$ vacua is dual to a CFT$_2$ with left- and right-moving central charges $c_L=0$ and $c_R=24k$, respectively, where $k$ is the level of the original Chern-Simons action. 

In addition to the Chiral Gravity sector, which corresponds to constant contorsion, the theory includes other interesting sectors. In particular, we presented an exact solution with non-constant curvature, asymptotically AdS$_3$ in the Brown-Henneaux sense. The theory admits the solutions of Chiral Gravity, such as the BTZ black holes. The values of the central charges of the conjectured CFT$_2$ agree with the those needed for the Cardy formula to reproduce the black holes entropy. 

The theory we studied here can be naturally extended to include a vector field. This can be achieved by decomposing the contortion in its irreducible parts $\kappa_{ab}=-\phi \epsilon_{abc}e^c - A_{[a}e_{b]} + M_{abc}e^c$, where the first term is the completely antisymmetric part, while the second and third terms correspond to the vector and the traceless symmetric parts, respectively. This may represent an interesting way of coupling Chiral Gravity to new matter fields. 

\begin{equation*}
\end{equation*}

The authors thank Hern\'an Gonz\'alez, Patricio Salgado and Ricardo Troncoso for discussions. The work of S.dP. was supported by program MECESUP 0806 and MECESUP CD FSM1204. The work of A.T. is supported by program MECESUP 0605 and MECESUP CD FSM1204. The work of G.G. was partially funded by FNRS-Belgium (convention FRFC PDR T.1025.14 and convention IISN 4.4503.15), by the Communaut\'{e} Fran\c{c}aise de Belgique through the ARC program and by a donation from the Solvay family. It was also supported by grants PIP0595/13 and UBACyT 20020120100154BA, from Consejo Nacional de Investigaciones Cient\'{\i}ficas y T\'{e}cnicas\ and Universidad de Buenos Aires. The work of J.Z. was partially supported by FONDECYT 1140155. The Centro de Estudios Cient\'{\i}ficos (CECs) is funded by the Chilean Government through the Centers of Excellence Base Financing Program of CONICYT- Chile.

\end{document}